\documentclass[aps,prl,twocolumn,superscriptaddress,showpacs]{revtex4}
\usepackage{epsfig}
\usepackage{graphicx}
\usepackage{dcolumn}
\usepackage{bm}
\usepackage{color}

\begin{document}

\title{Correlation-driven topological Fermi surface transition in FeSe}

\author{I. Leonov}
\affiliation{Theoretical Physics III, Center for Electronic Correlations and Magnetism, Institute of Physics, University of Augsburg, Augsburg 86135, Germany}
\author{S. L. Skornyakov}
\affiliation{Institute of Metal Physics, Sofia Kovalevskaya Street 18, 620990 Yekaterinburg GSP-170, Russia}
\affiliation{Ural Federal University, 620002 Yekaterinburg, Russia}
\author{V. I. Anisimov}
\affiliation{Institute of Metal Physics, Sofia Kovalevskaya Street 18, 620990 Yekaterinburg GSP-170, Russia}
\affiliation{Ural Federal University, 620002 Yekaterinburg, Russia}
\author{D. Vollhardt}
\affiliation{Theoretical Physics III, Center for Electronic Correlations and Magnetism, Institute of Physics, University of Augsburg, Augsburg 86135, Germany}

\date{\today}

\begin{abstract}

The electronic structure and phase stability of paramagnetic FeSe is computed by using a 
combination of {\it ab initio} methods for calculating band structure and dynamical mean-field 
theory. Our results reveal a topological change (Lifshitz transition) of the Fermi surface upon 
a moderate expansion of the lattice. The Lifshitz transition is accompanied with a sharp increase 
of the local moments and results in an entire reconstruction of magnetic correlations from the 
in-plane magnetic wave vector $(\pi,\pi)$ to $(\pi,0)$. 
We attribute this behavior to a correlation-induced shift of the Van Hove singularity originating 
from the $d_{xy}$ and $d_{xz}/d_{yz}$ bands at the M-point across the Fermi level.
We propose that superconductivity is strongly influenced, or even induced, by a Van Hove singularity.

\end{abstract}

 \maketitle

%%%%%%%%%%%%%%%%%%%%%%%%%%%%%%%%%%%%%%%%%%%%%
% Inroduction, previous LDA/LDA+DMFT results
%%%%%%%%%%%%%%%%%%%%%%%%%%%%%%%%%%%%%%%%%%%%%

The discovery of high-temperature superconductivity in iron pnictides \cite{KW08}, with critical temperatures 
$T_c$ up to 55 K, has lead to intensive experimental and theoretical research \cite{Review}. 
More recently, superconductivity has also been reported in the structurally related iron 
chalcogenide Fe$_{1+y}$Se close to its stoichiometric solution \cite{Hsu08}, with $T_c\sim 8$ K.
Structurally FeSe is the simplest of the Fe-based superconductors. It has the same layer structure as 
the iron pnictides, but without separating layers \cite{struct}.
Therefore FeSe is regarded as the parent compound for the Fe-based superconductors.
The critical temperature $T_c$ of FeSe depends very sensitively on changes of the lattice 
structure due to pressure or chemical doping. In particular, $T_c$ 
is found to increase up to $\sim 37$ K \cite{MT08,MM09} under hydrostatic pressure of $\sim 7$ GPa
and to $\sim 14$~K upon chemical (isovalent) substitution with Te \cite{SS09}.
%, which leads to the expansion of the lattice. 

The electronic structure of iron chalcogenides is also very similar to that of FeAs based superconductors,
according to both the angle-resolved photoemission \cite{XQ09,TG10,NS10} and band structure calculations \cite{SZ08}.
In particular, FeSe has the same Fermi surface topology as the pnictides. It is 
characterized by an in-plane magnetic nesting wave vector $Q_m=(\pi,\pi)$, consistent with 
s$^{\pm}$ pairing symmetry \cite{MS08}.
Moreover, both pnictides and chalcogenides display a strong enhancement of short-range spin 
fluctuations near $T_c$, with a resonance at $Q_m=(\pi,\pi)$ in the spin excitation spectra \cite{CG08}. 
These results suggest a common origin of superconductivity in pnictides 
and chalcogenides, e.g., due to spin fluctuations. However, in 
contrast to iron pnictides, FeSe shows no static magnetic order \cite{MM09,BQ09}. 
Moreover, the related (isoelectronic) compound FeTe exhibits no superconductivity and has a 
long-range $Q_m=(\pi,0)$ antiferromagnetic order \cite{BQ09}. 
In addition, FeTe exhibits a remarkable phase transition under pressure, from a tetragonal to 
a collapsed-tetragonal phase \cite{LC09}, with a simultaneous collapse of local moments, 
indicating that the solid solution Fe(Se,Te) is close to an electronic and/or lattice transition.

The iron chalcogenides FeSe$_{1-x}$Te$_x$ have been intensively investigated using photoemission and 
angle-resolved photoemission \cite{XQ09,TG10,NS10,YW09}, which reveal a significant narrowing of the Fe $3d$ 
bandwidth by a factor of $\sim 2$. In particular, a large enhancement of the quasiparticle mass
in the range of $\sim 3-20$ was reported \cite{TG10}, implying a crucial importance of electronic 
correlations. 
State-of-the-art methods for the calculation of the electronic structure of correlated 
electron materials, using the local-density approximation combined with dynamical mean-field 
theory (LDA+DMFT) approach \cite{DMFT,LDA+DMFT}, have shown to provide a good quantitative 
description of the electronic structure of iron pnictides and chalcogenides \cite{HS08,AB10}.
In particular, these calculations demonstrate the existence of a lower Hubbard band at 
about -1.5 eV below the Fermi level in FeSe \cite{AB10}. Moreover, these results 
show a significant orbital-dependent mass enhancement in the range of $2-5$.
%, implying a non-Fermi liquid state caused by the formation of local moments.
%
However, even today, in spite of intensive research, a microscopic explanation of the 
electronic properties and magnetism of iron chalcogenides is lacking. In particular, 
the interplay between electronic correlations and the lattice degrees of freedom in FeSe 
has remained essentially unexplored. We will address this problem in our investigation 
and thereby provide a microscopic explanation of the electronic structure and magnetic 
properties of the iron chalcogenide FeSe.

%%%%%%%%%%%%%%%%%%%%%%%%%%%%%%%%%%%%%%
% non-magnetic GGA results / GGA+DMFT
%%%%%%%%%%%%%%%%%%%%%%%%%%%%%%%%%%%%%%

\begin{figure}[b]
\centering \vspace{5.0mm} \includegraphics[width=0.8\linewidth]{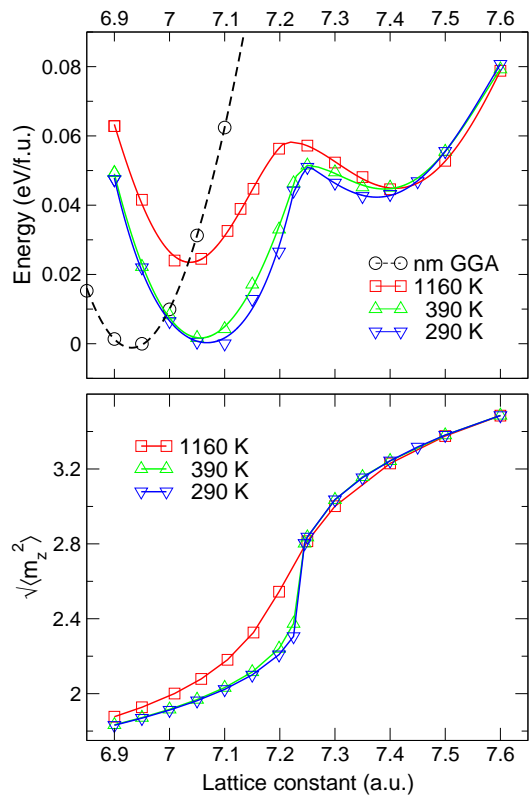}

\caption{(Color online) 
Total energy (upper panel) and mean fluctuating local moment (lower panel) of paramagnetic 
FeSe calculated for different temperatures by GGA+DMFT as a function of lattice constant. 
}

\label{Fig_1}
\end{figure}

In this Letter we employ the GGA+DMFT computational technique (GGA:
generalized gradient approximation) to explore the electronic structure and phase stability 
of the paramagnetic FeSe. In particular, we investigate the importance of electronic 
correlation effects for the electronic and magnetic properties of FeSe at finite temperatures. 
First we compute the electronic structure and phase stability of FeSe within the nonmagnetic 
GGA using the plane-wave pseudopotential approach \cite{Espresso}. To investigate the phase 
stability, we take a tetragonal crystal structure (space group $P4/mmm$) with the lattice 
parameter ratio $c/a = 1.458$ and Se position $z=0.266$ \cite{struct}, and calculate the total energy 
as a function of volume. 
Our results are presented in Fig.~\ref{Fig_1} (upper panel, dashed line); they are in good agreement 
with previous band-structure calculations \cite{SZ08}. The calculated equilibrium 
lattice constant is found to be $a \sim 6.92$ a.u., which is about $3$~\% lower than the 
experimental value \cite{struct}. The calculated bulk modulus is $B \sim $ 116 GPa \cite{bulk_modulus}.

To include the effect of electronic correlations, we employ the GGA+DMFT computational 
scheme. For the partially filled Fe $3d$ and Se $2p$ orbitals we construct a basis set 
of atomic-centered symmetry-constrained Wannier functions \cite{AK05}. To solve the realistic 
many-body problem, we employ the continuous-time hybridization-expansion quantum Monte-Carlo 
algorithm \cite{ctqmc,extra}. 
The calculations are performed at three different temperatures: 
$T = 290$~K, 390~K, and 1160~K. In these calculations we use the average Coulomb interaction 
${\bar U}=3.5$ eV and Hund's exchange $J=0.85$ eV, in accord with previous estimates for 
pnictides and chalcogenides \cite{HS08}. 
They are assumed to remain constant upon variation of the lattice volume. 
We employ the fully-localized double-counting correction, evaluated from the self-consistently 
determined local occupancies, to account for the electronic interactions already described 
by GGA.

%%%%%%%%%%%%%%%%%%%%%%%%%%%%%%%%%
% GGA+DMFT results: Total energy
%%%%%%%%%%%%%%%%%%%%%%%%%%%%%%%%%

In Fig.~\ref{Fig_1} (upper panel) we show the dependence of the total energy of paramagnetic 
FeSe as a function of lattice volume. Our result for the equilibrium lattice constant which 
now includes the effect of electronic correlations, agrees well with experiment. In particular, 
we find the equilibrium lattice constant $a = 7.07$~a.u., which is less than 1 \% off 
the experimental value. The calculated bulk modulus is $B \sim $ 70 GPa \cite{bulk_modulus}, 
which is comparable with that for iron pnictides \cite{bulk_modulus_examples}.
This is much lower than the result obtained without electronic correlations. Indeed,
the repulsive interaction leads to an increase of the unit cell volume 
and hence results in a reduction of the bulk modulus.
Most importantly, our result exhibits \emph{two} well-defined energy minima, one at 
$a \sim 7.1$ a.u. and another one at $a \sim 7.35$~a.u. Hence we predict a structural 
transition of FeSe upon a $\sim 10$\% expansion of the lattice volume corresponding to 
a negative pressure $p \sim -6.4$ GPa.
This result is unexpected and is very different from that obtained with the nonmagnetic GGA.
%
%The structural transition sets in upon an $\frac{\Delta V}{V} \sim 10$\% expansion of 
%the lattice, at critical pressure $p_c \sim -6.4$ GPa. 
At ambient pressure the high-volume tetragonal phase is only metastable, with a total energy 
difference w.r.t. to the equilibrium phase $\sim 42$~meV/f.u. at $T = 290$~K. 
The phase transition is of first order with an energy barrier of $\sim 10-15$~meV.
We interpret this behavior as a collapsed-tetragonal (low-volume) to tetragonal 
(high-volume) phase transformation upon expansion of the lattice volume. 
The phase transition is accompanied by a strong increase of the fluctuating local 
moment $\sqrt{ \langle m^2_z \rangle}$ [see Fig.~\ref{Fig_1} (bottom)], which grows 
monotonically upon expansion of the lattice.
The collapsed-tetragonal phase has a local moment $\sqrt{ \langle m^2_z \rangle} 
\sim 2$ $\mu_B$. By contrast, the high-volume phase has a much larger local moment 
of 3.25 $\mu_B$ and a softer lattice with a much lower bulk modulus of 35 GPa.
The existence of a second minimum in the total energy at a higher volume suggests the stability
of an isostructural compound with a larger lattice constant \cite{LC09}. This is indeed the case 
with FeTe, since the ionic radius of Te is larger than that of Se. Such an expansion of the 
lattice is known to increase $T_c$ by a factor of $\sim 2$, up to a maximum value $T_c \sim 14$ K \cite{SS09}.

%%%%%%%%%%%%%%%%%%%%%%%%%%%%%%%%%%%%%%%%%%%%%%%%%%%%%%%%%%%%%%%
% GGA+DMFT results: spectral function and mass renormalization
%%%%%%%%%%%%%%%%%%%%%%%%%%%%%%%%%%%%%%%%%%%%%%%%%%%%%%%%%%%%%%%

\begin{figure}[b]
\centering \vspace{5.0mm} \includegraphics[angle=-90]{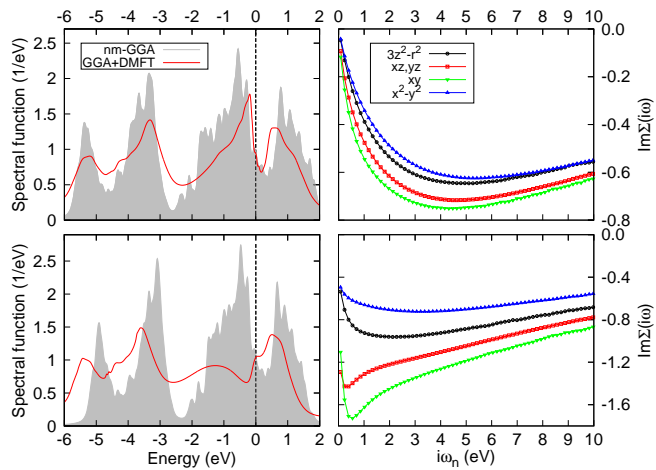}

\caption{(Color online) 
Left panels: spectral functions of paramagnetic FeSe obtained within nonmagnetic GGA
(shaded areas) and GGA+DMFT (straight lines). Right panels: orbitally-resolved imaginary 
parts of the self-energies as computed by GGA+DMFT. Upper row corresponds to the lattice 
constant $a = 7.1$ a.u., while the lower graphs display results for $a = 7.35$ a.u..
}
\label{Fig_2}
\end{figure}

Now we explore the origin of this surprising finding. For this purpose
we compute the total spectral function of paramagnetic FeSe using the GGA+DMFT 
approach. In Fig.~\ref{Fig_2} (top row) we display our results obtained for the 
collapsed-tetragonal phase with the lattice constant $a = 7.1$~a.u. The results 
for the high-volume tetragonal phase with $a = 7.35$~a.u. are shown in Fig.~\ref{Fig_2} (bottom row). 
In agreement with previous investigations \cite{AB10}, we find a reduction of 
the Fe $3d$ bandwidth near the Fermi energy caused by electronic correlations. 
The lower Hubbard band is located at about -1.5 eV for both phases. Upon 
expansion of the lattice, we observe a substantial spectral weight transfer, caused by 
strong electronic correlations. In particular, the spectral function for the low-volume phase exhibits 
a well-defined quasiparticle peak located below the Fermi level at $-0.19$ eV, which is absent in the 
high-volume phase. We note that the peak originates from the Van Hove 
singularity of the $d_{xz}/d_{yz}$ and $d_{xy}$ bands at the M-point. 

Our calculations reveal a remarkable orbital-selective renormalization of the Fe $3d$ 
bands, with significantly stronger correlations for the $d_{xz}/d_{yz}$ and $d_{xy}$, 
while the $d_{z^2}$ and $d_{x^2-y^2}$ bands exhibit weaker correlations. 
In the low-volume phase, the Fe $3d$ orbitals obey a Fermi-liquid like 
behavior with a weak damping at the Fermi energy.
The $d_{xz}$/$d_{yz}$ and $d_{xy}$ orbitals yield low effective quasiparticle 
weights $Z = (1 - \frac{\partial Im \Sigma(i \omega )}{\partial i\omega})^{-1}|_{\omega=0}$ 
of $\sim 0.48$ and $0.42$, respectively, while the self-energy for the $d_{x^2-y^2}$ 
and $d_{z^2}$ orbitals gives larger values of $0.65$ and $0.63$, respectively. Therefore
the quasiparticle mass enhancement is $\frac{m^*}{m} \sim 2.1$ for the $d_{xz}$/$d_{yz}$ 
and $\sim 2.4$ for the $d_{xy}$ orbitals, respectively.
In addition, we notice a substantial qualitative change in the self-energy 
upon expansion of the lattice. 
%
%The $d_{xz}$/$d_{yz}$ and $d_{xy}$ orbitals exhibit a clear trend to discontinuity in the high-volume phase. 
%
The calculated effective quasiparticle weights are $0.25$ and $0.17$.
Furthermore, the overall damping of quasiparticles becomes $\sim 6$ times larger, 
which implies a strong enhancement of electronic correlations.
For the high-volume phase, our calculations yield an effective quasiparticle mass 
enhancement of $\sim 4.0$ for the $d_{xz}$/$d_{yz}$ orbitals to $\sim 6.1$ for the $d_{xy}$.
These results show, in particular, that the effect of orbital-selective correlations increases 
upon expansion of the lattice.

%%%%%%%%%%%%%%%%%%%%%%%%%%%%%%%%%%%%%%%%%%%%%%%%%%%%%%%%%%%%%%%%
% GGA+DMFT result: fermi surface, topological phase transition
%%%%%%%%%%%%%%%%%%%%%%%%%%%%%%%%%%%%%%%%%%%%%%%%%%%%%%%%%%%%%%%%

\begin{figure}[b]
\centering \vspace{5.0mm} \includegraphics[angle=-90]{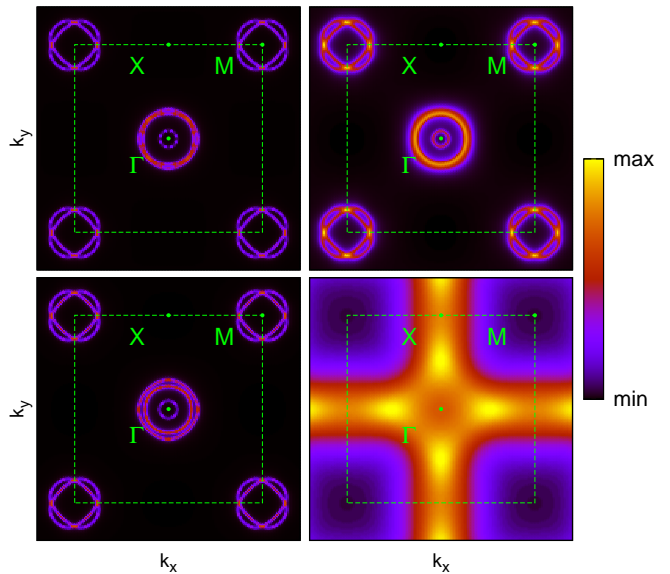}

\caption{(Color online) 
Fermi surface reconstruction in the $(\mathrm{k}_x,\mathrm{k}_y)$ plane at $\mathrm{k}_z=0$,
calculated for paramagnetic FeSe using nonmagnetic GGA (left panels) and GGA+DMFT (right 
panels) for the lattice constant $a = 7.1$ a.u. (upper row) and $a = 7.35$ a.u. (bottom row).
}
\label{Fig_3}
\end{figure}

Next we calculate the \textbf{k}-resolved spectra.
In Fig.~\ref{Fig_3} we display our results for the Fermi surface calculated for $k_z=0$. 
Again the nonmagnetic GGA results agree well with previous band-structure calculations \cite{SZ08}. 
We obtain two intersecting elliptical electron Fermi surfaces centered at the Brillouin 
zone M-point. In addition, there are three concentric hole pockets at the $\Gamma$-point 
(the two outer hole pockets are degenerate in the low-volume phase). 
In agreement with previous studies \cite{SZ08}, the Fermi surface topology shows 
the in-plane nesting with $Q_m=(\pi,\pi)$.
The nonmagnetic GGA calculations reveal no substantial change in the Fermi surface 
of FeSe upon expansion of the lattice. By contrast, the inclusion of correlation effects
leads to a complete reconstruction of the electronic structure upon expansion of the 
lattice \cite{GP13}, resulting in a dramatic change of the Fermi surface topology (Lifshitz transition).
In particular, the Fermi surface at the M-point collapses, leading to a large square-like 
hole pocket around the M-point in the high-volume phase, in surprising analogy with the cuprates.
In addition, the hole pockets around the $\Gamma$-point transform into incoherent 
spectral weight at the Fermi level along the $\Gamma$-X direction. 
The change of the Fermi surface topology results in a corresponding change of the magnetic 
correlations in FeSe. We find in-plane nesting with $Q_m=(\pi,\pi)$, connecting hole and 
electron parts of the Fermi surface, to be dominant in the low-volume phase. Upon expansion 
of the lattice by $\sim 5$ \%, i.e., at the energy maximum, the Lifshitz transition sets in, 
resulting in the $(\pi,0)$-type magnetic correlations in the high-volume phase.

%%%%%%%%%%%%%%%%%%%%%%%%%%%%%%%%%%%%%%%%%%%%%%%%%%%%%%%%%%%%%%%%%%%%%%%
% GGA+DMFT result: electronic structure, vH singularity at the M-point
%%%%%%%%%%%%%%%%%%%%%%%%%%%%%%%%%%%%%%%%%%%%%%%%%%%%%%%%%%%%%%%%%%%%%%%

\begin{figure}[b]
\centering \vspace{5.0mm} \includegraphics[angle=-90]{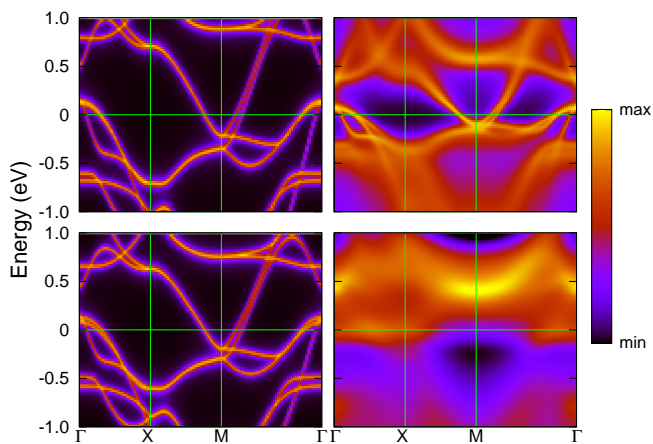}

\caption{(Color online) 
The {\bf k}-resolved spectral function of paramagnetic FeSe computed within nonmagnetic 
GGA (left panels) and GGA+DMFT (right panels) along the path $\Gamma$-X-M-$\Gamma$ 
for the lattice constant $a = 7.1$ a.u. (upper row) and $a = 7.35$ a.u. (bottom row).
}
\label{Fig_4}
\end{figure}

We have also calculated the momentum-resolved spectral functions along the high-symmetry 
directions (Fig.~\ref{Fig_4}). We find that a simple rescaling of the GGA band structure 
is not sufficient to account for the GGA+DMFT quasiparticle band structure, or the experimental data.
The effective crystal-field splitting between the Fe $3d$ orbitals is substantially 
renormalized because of the strong energy and orbital dependence of the self-energy, 
leading to different shifts of the quasiparticle bands near the Fermi level. 
In particular, we observe that the hole pockets near the $\Gamma$-point are pushed 
downward, while the states near the M-point are pushed upward, both towards the Fermi 
level [see Fig.~\ref{Fig_4} (upper row)], in agreement with the ARPES measurements 
\cite{TG10,NS10}. This indicates that charge transfer caused by electronic correlations 
is important, resulting in a substantial shift of the Van Hove singularity at the M-point 
towards the Fermi level, while the nonmagnetic GGA band structure depends only weakly on 
an expansion of the lattice. The GGA+DMFT results show an entire reconstruction
of the electronic structure of paramagnetic FeSe in the high-volume phase. This behavior 
is found to be associated with a correlation-induced shift of the Van Hove singularity in 
the M-point \emph{above} the Fermi level. It results in a non-Fermi-liquid like behavior 
and strong enhancement of the effective electron mass at the phase transition.

%%%%%%%%%%%%%%%%%%%%%%%%%%%%%
% short discussion
%%%%%%%%%%%%%%%%%%%%%%%%%%%%%

Our results indicate the crucial importance of the proximity of a Van Hove singularity to 
the Fermi level for the appearance of unconventional superconductivity in the chalcogenide 
FeSe$_{1-x}$Te$_x$ series. 
Indeed, we propose that the superconductivity is strongly influenced, or even induced, by a Van Hove 
singularity.
Furthermore, we predict a topological change (Lifshitz transition) of the Fermi surface on doping FeSe 
by Te, which is accompanied with a sharp increase of the local moments \cite{c_a}.
We further expect that these changes are responsible for the experimentally observed increase of 
$T_c$ in FeSe upon doping with Te. The microscopic origin for superconductivity would then be a
Van Hove singularity close to the Fermi level in this system \cite{cuprates}. This identification 
may open a new route to increase $T_c$ even further.

In conclusion, we employed the GGA+DMFT computational technique to explore the 
electronic structure and phase stability of the paramagnetic tetragonal phase 
of FeSe. Our results clearly demonstrate the crucial importance of electronic 
correlations on the properties of FeSe at finite temperatures. 
In particular, they reveal a complete reconstruction of the Fermi surface topology 
upon a moderate expansion of the lattice, which is accompanied with a change of 
magnetic correlations from the in-plane magnetic wave vector $(\pi,\pi)$ to $(\pi,0)$.
We attribute this behavior to the formation of local moments which are caused by a 
correlation-induced shift of the Van Hove singularity. The latter originates from the $d_{xy}$ 
and $d_{xz}$/$d_{yz}$ bands at the M-point across the Fermi level. 
In addition, we observe an orbital-dependent renormalization of the Fe $3d$ bands 
near the Fermi level, where the $d_{xy}$ bands are heavily renormalized compared 
to the $d_{xz}$/$d_{yz}$ orbitals.
Our results suggest that the proximity of the Van Hove singularity to the Fermi level 
is responsible for the unconventional superconductivity in the chalcogenide FeSe$_{1-x}$Te$_x$ series.

\begin{acknowledgments}

We thank Vladimir Tsurkan for valuable discussions.
The authors acknowledge support of the Russian Scientific Foundation (project no. 14-22-00004), 
the Russian Foundation for Basic Research (projects no. 13-02-00050, no. 13-03-00641), the Ural 
Division of the Russian Academy of Science Presidium (project no. 14-2-NP-164, no. 12-P2-1017) 
(S.L.S. and V.I.A.), and the Deutsche Forschergemeinschaft through TRR 80 (I.L.) and FOR 1346 (D.V.). 
S.L.S. is grateful to the Dynasty Foundation.

\end{acknowledgments}


\begin{thebibliography}{99}
\bibitem{bulk_modulus}
The bulk modulus is calculated as a second derivative of the spline interpolation of $E(V)$, $B=V \frac{\partial^2 E}{\partial V^2}$.
\bibitem{extra}
We here employ the density-density approximation of the local Coulomb interaction between the Fe $3d$ electrons. 
Our results shown in Fig.~\ref{Fig_2} agree well with those obtained with the full rotationally invariant 
Hund's coupling \cite{AB10}. We notice only a minor multiplet structure below the Fermi level missing in 
our calculation.
\bibitem{c_a}
We also note that similar behavior is found to occur upon an uniaxial compression of paramagnetic FeSe 
along the $c$-axis (not shown). Our preliminary results suggest the same microscopic origin,
namely, the proximity of a Van Hove singularity to the Fermi level, to be responsible for the experimentally 
observed increase of $T_c$ in FeSe under pressure.
\end{thebibliography}
\end{document}